\documentclass{emulateapj}
\usepackage{apjfonts}
\usepackage{graphicx}

\shortauthors{Winn, Fabrycky, Albrecht, Johnson 2010}
\shorttitle{Hot Stars}

\begin{document}

%
\def\ltsima{$\; \buildrel < \over \sim \;$}
\def\lsim{\lower.5ex\hbox{\ltsima}}
\def\gtsima{$\; \buildrel > \over \sim \;$}
\def\gsim{\lower.5ex\hbox{\gtsima}}
                                                                                          
%

\bibliographystyle{apj}

\title{ Hot Stars with Hot Jupiters Have High Obliquities }

\author{
Joshua N.\ Winn\altaffilmark{1},
Daniel Fabrycky\altaffilmark{2,3},
Simon Albrecht\altaffilmark{1},
John Asher Johnson\altaffilmark{4}
}

\journalinfo{Draft version, June 21, 2010}
\slugcomment{Submitted to ApJ Letters, May 3, 2010} 

\altaffiltext{1}{Department of Physics, and Kavli Institute for
  Astrophysics and Space Research, Massachusetts Institute of
  Technology, Cambridge, MA 02139}

\altaffiltext{2}{Harvard-Smithsonian Center for Astrophysics, 60
  Garden St., Cambridge, MA 02138}

\altaffiltext{3}{Michelson Fellow}

\altaffiltext{4}{Department of Astrophysics, and NASA Exoplanet
  Science Institute, California Institute of Technology, MC~249-17,
  Pasadena, CA 91125}

\begin{abstract}

  We show that stars with transiting planets for which the stellar
  obliquity is large are preferentially hot ($T_{\rm
    eff}~>~6250$~K). This could explain why small obliquities were
  observed in the earliest measurements, which focused on relatively
  cool stars drawn from Doppler surveys, as opposed to hotter stars
  that emerged later from transit surveys. The observed trend could be
  due to differences in planet formation and migration around stars of
  varying mass. Alternatively, we speculate that hot-Jupiter systems
  begin with a wide range of obliquities, but the photospheres of cool
  stars realign with the orbits due to tidal dissipation in their
  convective zones, while hot stars cannot realign because of their
  thinner convective zones. This in turn would suggest that hot
  Jupiters originate from few-body gravitational dynamics, and that
  disk migration plays at most a supporting role.

\end{abstract}


\section{Introduction}
\label{sec:introduction}

There are now 28 cases of stars with transiting planets for which the
stellar obliquity---or more precisely its sky projection---has been
measured via the Rossiter-McLaughlin effect. The history of these
measurements is perplexing. Starting with the pioneering measurement
of Queloz et al.~(2000), for 8 years a case was gradually building
that the orbits of hot Jupiters are always well-aligned with the
rotation of their parent stars. Then in a sudden reversal, several
misaligned systems were found, with the first sighting by H\'ebrard et
al.~(2008) and the most recent spate of discoveries by Triaud et
al.~(2010).

In this Letter we point out that the misaligned systems are
preferentially those with the hottest photospheres. In \S~2 we discuss
the sample, and in \S~3 we display the patterns involving the order in
which the measurements were made, the stellar effective temperature,
and the stellar obliquity. In \S~4 we speculate on the meaning of the
patterns, and in \S~5 we summarize the results and their implications
for theories of the origin of hot Jupiters.

\section{The Sample}
\label{sec:sample}

We focused on those systems for which the projected spin-orbit angle,
$\lambda$, was measured with a 1$\sigma$ precision of $10^\circ$ or
better. The less precise cases are not as helpful because we cannot
tell definitively whether the system is aligned or misaligned, and
because the large uncertainties are usually associated with strong
systematic effects.

We omitted {\it Kepler}-8 (Jenkins et al.~2010) from consideration
even though the quoted uncertainty is smaller than $10\arcdeg$,
because no data were gathered immediately before or after the transit,
precluding tests for a systematic velocity offset on the transit
night. Such offsets are possible, or even probable, for stars as faint
as {\it Kepler}-8 observed in bright moonlight (see, e.g., Tripathi et
al.~2010). When we reanalyzed the {\it Kepler}-8 data allowing for
such an offset, the result was $\lambda = 20^\circ \pm 20^\circ$.

Table~1 summarizes the resulting sample of 19 systems, along with the
properties of the 9 omitted systems, for completeness. For simplicity
we refer to the planets as ``hot Jupiters'' because they are all giant
planets with short periastron distances, although it should be
remembered that they span a wide range of masses (0.36--11.8~$M_{\rm
  Jup}$) and orbital periods (1.3--111~d).

\section{The Pattern}
\label{sec:pattern}

\begin{figure*}[ht]
\begin{center}
 \leavevmode
\hbox{
 \epsfxsize=5in
 \epsffile{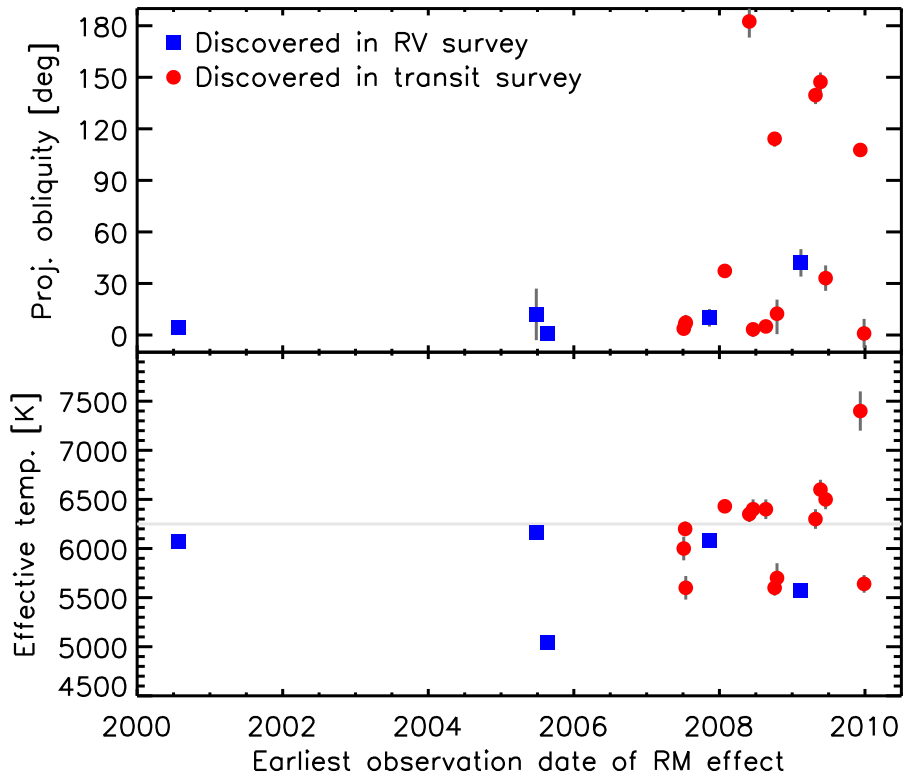}}
\end{center}
\caption{{\bf A brief history of Rossiter-McLaughlin measurements.}
{\it Top.}---The projected obliquity (spin-orbit angle) is plotted as a function
of the earliest date of observation reported in the literature. Blue
squares indicate systems discovered by RV surveys,
while red circles indicate systems
found in photometric transit surveys.
{\it Bottom.}---The
stellar effective temperature of the same systems.
Systems with $T_{\rm eff}$ higher than 6250~K have been
discovered by transit surveys, and began to be examined for the RM
effect in 2008.
\label{fig:time}}
\end{figure*}

The top panel of Fig.~1 shows $\lambda$ as a function of the date of
the earliest reported observation of the Rossiter-McLaughlin
effect. The trend of low values (good alignment) for the first few
years is evident, as is the ``spike'' of high values (misalignment) in
the most recent years. This plot also suggests that the systems
initially discovered in radial-velocity (RV) surveys are
systematically more well-aligned than those systems discovered in
transit surveys.

The reason for this pattern may be that the earlier measurements
focused on cooler and less massive stars. The bottom panel of Fig.~1
shows that the average effective temperature ($T_{\rm eff}$) of the
host stars has risen with time. Only in 2008 did investigators begin
examining stars with $T_{\rm eff} > 6250$~K, and all of those systems
were identified in transit surveys as opposed to RV surveys.

We cannot give a deterministic explanation for this trend, as it
depends not only on the selection functions for the various surveys
but also sociological factors affecting the allocation of telescope
time. However it seems probable that cooler stars were examined
earlier because they allow for better RV precision, and therefore
greater ease of confirming the existence of planets. Indeed, most RV
surveys exclude early-type stars altogether. In contrast, transit
surveys have nearly magnitude-limited samples that include hot and
luminous stars. These factors may explain why planets around hot stars
were only found in transit surveys, and why they emerged relatively
late from those surveys.

Fig.~2 shows $\lambda$ as a function of $T_{\rm eff}$. Most of the
misaligned systems are around the hottest stars in the sample. The
transition from aligned to misaligned occurs around $T_{\rm eff} =
6250$~K (spectral type F8), which for the rest of this Letter we take
to be the boundary between ``cool'' and ``hot'' stars. We will also
use the term ``misaligned'' to mean $|\lambda|>10^\circ$ with
$>$3$\sigma$ confidence.

Among the cool stars, two out of 11 (18\%) are misaligned, while among
the hot stars, 6 out of 8 (75\%) are misaligned. Another way to
describe the pattern is to enumerate exceptions to the rule that only
hot stars are misaligned. There are two types of exceptions:
``strong'' exceptions in which a cool star is misaligned, and ``weak''
exceptions in which a hot star is apparently well-aligned. Weak
exceptions are not as serious because only the sky-projected obliquity
is measured, and consequently a low value of $\lambda$ could be
observed for a misaligned system. Out of 19 systems, there are two
strong exceptions and two weak exceptions.

\begin{figure*}[ht]
\begin{center}
  \leavevmode
\hbox{
  \epsfxsize=5in
  \epsffile{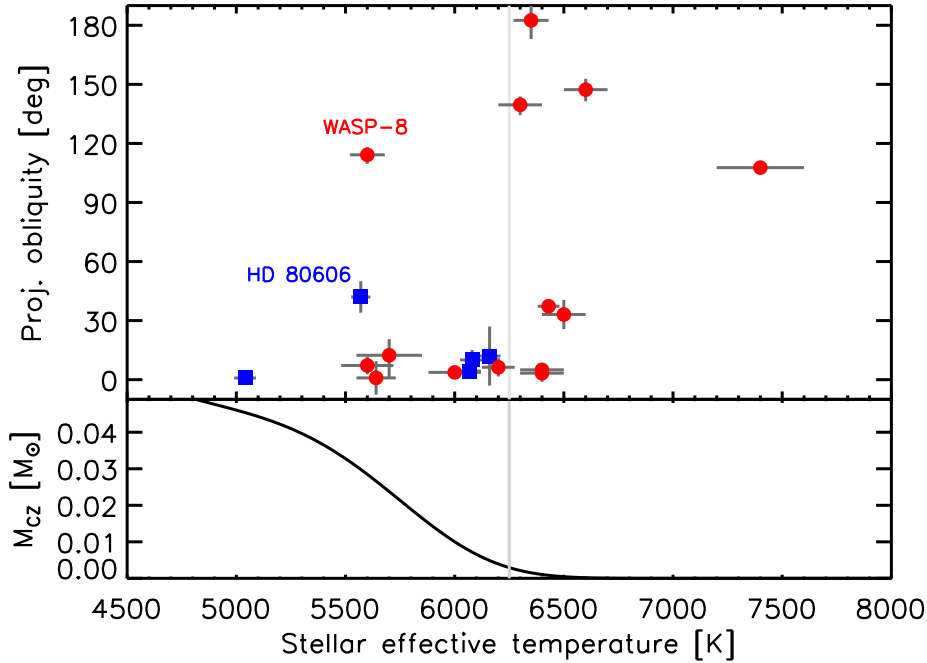}}
\end{center}
\caption{ {\bf Misaligned systems have hotter stars.}  {\it
    Top.}---The projected obliquity is plotted against the effective
  temperature of the host star. A transition from mainly-aligned to
  mainly-misaligned seems to occur at $T_{\rm eff} \approx
  6250$~K. The two strongest exceptions are labeled. Symbol colors and
  shapes have the same meaning as in Fig.~1. {\it Bottom.}---The mass
  of the convective zone of a main-sequence star as a function of
  $T_{\rm eff}$, from Pinsonneault, DePoy, \& Coffee~(2001). It is
  suggestive that 6250~K is approximately the temperature at which the
  mass of the convective zone has bottomed
  out.\label{fig:temp_lambda}}
\end{figure*}

Many seemingly compelling trends of this kind turn out to be
spurious. The best way to make progress is to gather more data. The
prediction that misaligned systems are preferentially around hot stars
will be tested in the near future, and a primary purpose of this
Letter is to enunciate the prediction in advance of forthcoming
observations.

It is also important to consider selection effects. From a transit
surveyor's perspective, the most important difference between a
well-aligned star and a misaligned star of the same spectral type is
that the well-aligned star has a larger $v\sin i$ (sky-projected
rotation rate). A large $v\sin i$ implies broader spectral lines and
poorer Doppler precision, inhibiting planet discovery. Therefore there
is a potential bias against discovering well-aligned systems. We must
ask whether there could exist a large population of well-aligned
systems around hot stars that has been missed by current surveys.

For this question, the RV surveys are irrelevant because they exclude
all hot stars regardless of $v\sin i$. As for the transit surveys, to
assess the bias we must know how transit candidates are identified and
followed up. Latham et al.~(2009) provided a complete inventory of
transit candidates and follow-up observations, which we take to be
representative. They chose 28 transit candidates for spectroscopic
follow-up, without regard to spectral type. Of those, 4 were not
pursued further once it was found that $v\sin i>50$~km~s$^{-1}$. The
other 24 cases were observed assiduously until a hot Jupiter was
confirmed (2 cases) or ruled out (22 cases). Hence, any bias against
well-aligned systems is probably only for stars with $v\sin i \gsim
50$~km~s$^{-1}$. Such rapid rotators typically have spectral types
$<$F3 and $T_{\rm eff}>6700$~K, whereas our sample ranges from 5040 to
6700~K, with one exception.\footnote{The exception, WASP-33 (7200~K),
  proves the rule. The planet was discovered despite the star's rapid
  rotation ($v\sin i=86$~km~s$^{-1}$) by exploiting the RM effect
  (Collier Cameron et al.~2010) and not by the usual procedure of
  measuring the spectroscopic orbit. Therefore, even if the other hot
  systems were selected in a manner biased against well-aligned
  systems, we would expect WASP-33 to be more representative of the
  true obliquities of hot stars---and it is misaligned.}  Hence it
seems unlikely that this bias is completely responsible for the
observed $\lambda$-$T_{\rm eff}$ relation, although it may play some
role.

Another caveat is that we cannot tell whether the relevant parameter
is really $T_{\rm eff}$ or some other correlated variable, such as
stellar mass. The transition temperature of 6250~K corresponds to
approximately 1.3~$M_\odot$ for solar-metallicity main-sequence stars.

\section{ Possible Explanations }
\label{sec:explanation}

Despite these caveats it is impossible to resist speculating on the
reasons why hot stars with hot Jupiters have high obliquities. We
restrict ourselves to an airing of issues and a toy model illustrating
a speculative hypothesis, leaving detailed investigations for future
work.

One possibility is that there are two pathways for producing hot
Jupiters, one of which is specific to low-mass stars and yields low
obliquities, while the other occurs mainly for massive stars and
produces a broad range of obliquities. The low-obliquity mechanism
could be inspiral due to tidal interactions with the protoplanetary
disk (Lin, Bodenheimer, \& Richardson~1996). The high-obliquity
mechanism could be some combination of planet-planet scattering
(Chatterjee et al.~2008) and Kozai cycles (Fabrycky \& Tremaine
2007). It is not obvious why these mechanisms would have a strong
dependence on stellar mass or temperature, although it is interesting
that 1.3~$M_\odot$ is approximately the same stellar mass above which
giant planets are found to have larger masses, wider orbits, and a
higher rate of occurrence (Bowler et al.~2010). Perhaps more massive
stars are more likely to form systems of massive planets in unstable
configurations, leading to an enhanced rate of gravitational
scattering in comparison to cooler stars.

Another possibility is suggested by the sharpness of the transition
from aligned to misaligned, and its location at $T_{\rm eff} \approx
6250$~K. For main-sequence stars, this is approximately the
temperature above which the mass in the outer convective zone ($M_{\rm
  cz}$) becomes inconsequential. The decline in $M_{\rm cz}$ is
illustrated in the bottom panel of Fig.~2, based on the relation
presented by Pinsonneault, DePoy, \& Coffee~(2001). Between spectral
types G0 and F5 (5940 and 6650~K), $M_\star$ increases by a factor of
1.3, and $M_{\rm cz}$ decreases by a factor of 120.

Convective zones are important for the production of magnetic fields
and for tidal dissipation. Magnetic fields may be relevant by setting
the inner radius of the protoplanetary gas disk, where accreting
material is captured onto field lines, or by allowing the star to spin
down through magnetic braking. The possible relevance of tidal
dissipation is even more obvious, as it would tend to realign the star
with the orbit.

Pursuing this latter point, we hypothesize that there is a single
mechanism for producing hot Jupiters, and this mechanism yields a
broad range of obliquities. For the cool stars, tidal dissipation
damps the obliquity within a few Gyr, while for the hot stars,
dissipation is ineffective. Therefore we observe hot Jupiters to be
well-aligned around cool stars, and misaligned around hot stars.

It has been argued previously that tidal dissipation is too slow to
affect the stellar spin state (see, e.g., Winn et al.~2005), but these
arguments should now be reconsidered. The timescales for tidal
dissipation are not understood from first principles and are poorly
constrained by observations. Another objection is that obliquity
damping should be accompanied by spin-orbit synchronization, which is
not observed. However, cool stars spin down due to magnetic
braking. Thus, even if tides do synchronize the rotation and orbital
periods while damping the obliquity, magnetic braking could
subsequently slow the rotation to the observed values. A third
objection, and the hardest to overcome, is that obliquity damping is
accompanied by orbital decay, threatening the planet with engulfment
(Levrard, Winisdoerffer, \& Chabrier 2009, Barker \& Ogilvie
2009). The planet must surrender all its angular momentum in order to
reorient the star, because of the star's large moment of inertia.

We are thereby led to explore a scenario in which the star's moment of
inertia is drastically reduced. We suppose that only the convective
zone is dissipatively torqued by the planet, and that the radiative
zone is weakly coupled to the convective zone and to the planet.
Without the burden of the massive radiative interior, the convective
zone---and thus the observable photosphere---can align with the
planetary orbit without drawing in the planet. Likewise, the magnetic
braking torque would be even more effective in slowing the surface
rotation speed and preventing spin-orbit synchronization.

Core-envelope decoupling has been discussed in the context of young
stars (see, e.g., Irwin \& Bouvier 2009), but here we would need
decoupling to persist for a sizable fraction of the main-sequence
lifetime of a cool star. A problem with this notion is that the Sun's
convective and radiative zones appear to be well-coupled (Howe
2009). However, this may not have always been so, and it was not a
foregone conclusion theoretically (see, e.g., Pinsonneault et
al.~1989). The most plausible solar coupling mechanisms, magnetic
linkage and internal gravity waves, may be absent or may act on longer
timescales for stars with hot Jupiters.

To investigate the effects of core-envelope decoupling we used the
equations of Eggleton \& Kiseleva-Eggleton (2001) to follow a circular
orbit of a hot Jupiter around a 1~$M_\odot$ star, with initial periods
$P_{\rm orb}=3$~d and $P_{\rm rot}=10$~d. Based on the stellar
evolution code
EZ-Web\footnote{http://www.astro.wisc.edu/$\sim$townsend/static.php?ref=ez-web}
we take the convective zone to have mass $0.015~M_\odot$, moment of
inertia $0.0066$~$M_\odot R_\odot^2$, and apsidal motion constant
$9\times10^{-4}$. We chose a tidal dissipation factor
$Q'_\star=6\times10^6$, which is consistent with the current
population of hot Jupiters, although the large uncertainty in
$Q'_\star$ causes a correspondingly large uncertainty in all of the
timescales reported here. We do not model the dissipative shear or the
non-dissipative oblateness coupling between the convective zone and
the radiative interior. The magnetic braking torque was modeled with
an extra term in the equations of motion:
\begin{equation}
\frac{d\vec{\Omega}_\star}{dt} = - \alpha_{\rm mb} \Omega_\star^2 \vec{\Omega}_\star.
\end{equation}
For the braking coefficient $\alpha_{\rm mb}$ we used
$1.66\times10^{-13}$~yr, based on a scaling of the Barker \& Ogilvie
(2009) results according to the moment of inertia.

Fig.~3 shows the time history of $P_{\rm rot}$, $P_{\rm orb}$, and the
stellar obliquity $\psi$, assuming an initial value of
60~deg. (Similar results were obtained from an initially retrograde
condition.) Three lines are plotted, corresponding to planet masses of
3, 1, or 1/3~$M_{\rm Jup}$. For Jupiter-mass planets, the stellar
obliquity damps before the planet is consumed. Magnetic braking
prevents synchonization of the convective zone with the orbit, in
agreement with observations. However, this model also implies that
orbits decay within main-sequence lifetimes, and that close-in massive
planets should be rarer around cool stars than hot stars, due to their
more rapid orbital decay.

Another prediction is that the planets exerting the weakest tidal
torques should be seen as ``strong exceptions'': misaligned planets
around cool stars. To compute the obliquity-damping component of the
tidal torque we averaged together the last terms of Eqns.~(10) and
(11) of Eggleton \& Kiseleva-Eggleton (2001), giving a decay timescale
proportional to
\begin{equation}
\left(\frac{M_{\rm cz}}{M_p}\right)
\left(\frac{a}{R_\star}\right)^6
\frac{(1-e^2)^{9/2}}{1+ 3e^2 + (3/8)e^4},
\end{equation}
where $M_p$ is the planet mass, $a$ is the orbital distance, $R_\star$
is the stellar radius, and $e$ is the orbital eccentricity.  By this
standard, the 3 systems with the longest timescales for obliquity
damping are HD~80606, HD~17156 and WASP-8. Thus, in our theory it is
appropriate that HD~80606 and WASP-8 are strong exceptions.

\begin{figure}[htb]
\epsscale{1.00}
\plotone{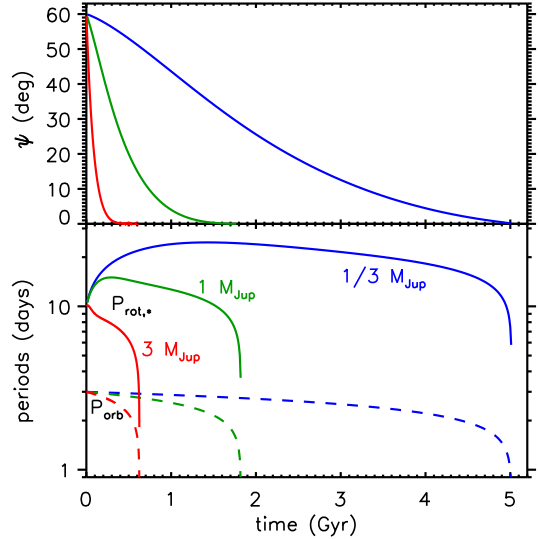}
\caption{ {\bf Toy model in which obliquity damping precedes
    orbital decay.} See the text for details. Each panel shows
  curves corresponding to planetary masses of 3, 1, and 1/3~$M_{\rm
    Jup}$.  {\it Top.}---The obliquity $\psi$ of the observable
  photosphere damps quickly, because the outer convective zone has
  little mass.  {\it Bottom.}---The star's rotation period, $P_{\rm
    rot,*}$ (solid lines), is kept slow by the action of magnetic braking. The
  planet's orbital period, $P_{\rm orb}$ (dashed lines), ultimately shrinks to zero
  (and the planet is destroyed), but on a longer timescale than
  obliquity damping.
  \label{fig:model} }
\end{figure}

\section{ Discussion }
\label{sec:discussion}

The finding that hot stars with hot Jupiters tend to have high
obliquities is not the only pattern that has been described in the
Rossiter-McLaughlin data. Johnson et al.~(2009) and H\'ebrard et
al.~(2010) found that the first 3 known misaligned systems all
involved relatively massive planets on eccentric orbits. Since then,
several exceptions have been discovered, such as WASP-15 and WASP-17
(Triaud et al.~2010).

The $\lambda$--$T_{\rm eff}$ relation may be a sign that the
mechanisms that produce hot Jupiters depend strongly on stellar
mass. We have also explored a theory in which hot Jupiters are
emplaced with a wide range of obliquities around all stars, but the
cool stars tidally realign with the planetary orbits. The main
difficulty with any theory of tidal realignment is avoiding orbital
decay. Core-envelope decoupling could postpone orbital decay until
after alignment is achieved, although this scenario is admittedly
speculative. One implication would be that close-in massive planets
should be rarer around cool stars. Another implication would be that
attempts to compare the ensemble results for $\lambda$ and the
predictions of migration theories, such as those of Fabrycky \& Winn
(2009) and Triaud et al.~(2010), should consider only hot stars,
because cool stars may have been affected by subsequent tidal
evolution.

Finally, we interpret the results, as did Triaud et al.~(2010), as a
blow against the theory of disk migration, which would yield low
obliquities as a general rule. Disk migration probably does play a
role in sculpting exoplanetary orbits, and convergent migration of
multiple planets may occasionally produce tilted orbits (Yu \&
Tremaine 2001). But if obliquity truly depends on the present-day
convective zone of the host star, then hot Jupiters likely arrived
after the pre-main sequence convective phase ceased, tens of Myr after
disk dispersal. Few-body gravitational dynamics (scattering or Kozai
cycles) followed by tidal dissipation in the planet is compatible with
this timescale, and it naturally produce misalignments, so this
mechanism might account for most or all hot Jupiters.

\acknowledgments We thank Jeremy Goodman, Matija Cuk, Bill Cochran,
Sally Dodson-Robinson, Andrew Howard, Geoff Marcy, Tim Morton, Fred
Rasio, Dimitar Sasselov, and Richard Wade for helpful
discussions. J.N.W.\ gratefully acknowledges support from the NASA
Origins program through award NNX09AD36G and the MIT Class of 1942, as
well as the Tinsley Scholars program and the hospitality of the
Astronomy Department at the University of Texas, Austin. DF
acknowledges support from the Michelson Fellowship, supported by the
National Aeronautics and Space Administration and administered by the
Michelson Science Center. SA acknowledges support from a NWO
Rubicon fellowship.\\

\begin{deluxetable}{lccccc}
 
\tabletypesize{\tiny}
\tablecaption{A brief history of Rossiter-McLaughlin measurements\label{tbl:sample}}
\tablewidth{0pt}
  
\tablehead{
\colhead{Name} &
\colhead{Earliest observation date} &
\colhead{Type of survey} &
\colhead{$T_{\rm eff}$~[K]} &
\colhead{$\lambda$~[deg]} &
\colhead{References}
}

\startdata
\\
HD~209458     & 2000~Jul~29 & RV      & $6070\pm 50$ & $-4.4\pm 1.4$ & 1,2 \\
HD~149026$^\star$ & 2005~Jun~26 & RV    & $6160\pm 50$ & $-12\pm 15$ & 3 \\
HD~189733     & 2005~Aug~21 & RV      & $5040\pm 50$ & $-0.85_{-0.32}^{+0.28}$ & 4,5 \\
TrES-1$^\star$  & 2006~Jun~21 & Transit & $5230\pm 50$ & $30\pm 21$ & 6 \\
TrES-2$^\star$  & 2007~Apr~26 & Transit & $5850\pm 50$ & $-9\pm 12$ & 7 \\
HAT-P-2$^\star$ & 2007~Jun~06 & Transit & $6290\pm 60$ & $0.2^{+12.2}_{-12.5}$ & 8,9 \\
HAT-P-1       & 2007~Jul~06 & Transit & $6000\pm 120$ & $3.7\pm 2.1$ & 10 \\
Corot-2       & 2007~Jul~16 & Transit & $5600\pm 120$ & $-7.2\pm 4.5$ & 11 \\
TrES-4        & 2007~Jul~13 & Transit & $6200\pm 75$ & $-6.3\pm 4.7$ & 12 \\
HD~17156      & 2007~Nov~12 & RV      & $6080\pm 56$ & $10.0\pm 5.1$ & 13,14,15,16 \\
XO-3          & 2008~Jan~28 & Transit & $6430\pm 50$ & $-37.3\pm 3.7$ & 17,18 \\
Corot-1$^\star$ & 2008~Feb~27 & Transit & $6000\pm 150$ & $77\pm 11$ & 19 \\
HAT-P-7       & 2008~May~30 & Transit & $6350\pm 80$ & $182.5\pm 9.4$ & 20,21 \\
WASP-3        & 2008~Jun~18 & Transit & $6400\pm 100$ & $3.3_{-4.4}^{+2.5}$ & 22,23 \\
WASP-18       & 2008~Aug~21 & Transit & $6400\pm 100$ & $-5.0_{-2.8}^{+3.1}$ & 24 \\
Corot-3$^\star$ & 2008~Aug~26 & Transit & $6700\pm 140$ & $-37.6_{-10.0}^{+22.3}$ & 25 \\
WASP-8        & 2008~Oct~04 & Transit & $5600\pm 80$ & $-114.2_{-4.6}^{+3.9}$ & 26 \\
WASP-4        & 2008~Oct~08 & Transit & $5500\pm 150$ & $4_{-43}^{+34}$ & 24 \\
WASP-6        & 2008~Oct~08 & Transit & $5500\pm 100$ & $-11_{-14}^{+18}$ & 27 \\
WASP-2$^\star$  & 2008~Oct~15 & Transit & $5200\pm 200$ & $-153_{-11}^{+15}$ & 24 \\
WASP-5        & 2008~Oct~16 & Transit & $5700\pm 150$ & $12.4_{-11.9}^{+8.2}$ & 24 \\
WASP-15       & 2009~Apr~27 & Transit & $6300\pm 100$ & $-139.6_{-5.2}^{+4.3}$ & 24 \\
WASP-17       & 2009~May~22 & Transit & $6600\pm 100$ & $-147.3_{-5.9}^{+5.5}$ & 24,28 \\
HD~80606      & 2009~Feb~13 & RV      & $5570\pm 44$ & $42\pm 8$ & 29,30,31,32 \\
WASP-14       & 2009~Jun~17 & Transit & $6500\pm 100$ & $-33.1\pm 7.4$ & 33 \\
{\it Kepler}-8$^\star$      & 2009~Oct~29 & Transit & $6200\pm 150$ & $-26.9\pm 4.6$ & 34 \\
WASP-33       & 2009~Dec~08 & Transit & $7400\pm 200$ & $-107.7\pm 1.6$ & 35 \\
HAT-P-13      & 2009~Dec~27 & Transit & $5640\pm 90$ & $-0.9\pm 8.5$ & 36\\
\enddata

\tablecomments{References:
(1) Winn et al.~(2005), (2) Queloz et al.~(2000),
(3) Wolf et al.~(2007),
(4) Triaud et al.~(2009), (5) Winn et al.~(2006),
(6) Narita et al.~(2007),
(7) Winn et al.~(2008),
(8) Loeillet et al.~(2008), (9) Winn et al.~(2007)
(10) Johnson et al.~(2008),
(11) Bouchy et al.~(2008),
(12) Narita et al.~(2010),
(13) Narita et al.~(2009a), (14) Barbieri et al.~(2009), (15) Cochran et al.~(2008), (16) Narita et al.~(2008),
(17) Winn et al.~(2009a), (18) H\'ebrard et al.~(2008),
(19) Pont et al.~(2010),
(20) Winn et al.~(2009c), (21) Narita et al.~(2009b),
(22) Tripathi et al.~(2010), (23) Simpson et al.~(2010),
(24) Triaud et al.~(2010),
(25) Triaud et al.~(2009),
(26) Queloz et al.~(2010),
(27) Gillon et al.~(2009),
(28) Anderson et al.~(2010),
(29) H\'ebrard et al.~(2010), (30) Moutou et al.~(2009), (31) Winn et al.~(2009b), (32) Pont et al.~(2010),
(33) Johnson et al.~(2009),
(34) Jenkins et al.~(2010),
(35) Collier Cameron et al.~(2010),
(36) Winn et al.~(2010).
Where more than one reference is given, the quoted value for $\lambda$
is taken from the first reference in the list. Some authors use a different
coordinate system and report $\beta\equiv -\lambda$; for this table
we have converted all results to $\lambda$.
For WASP-33 the tabulated value and error bar for $\lambda$ represent the mean
and standard deviation of the 3 independently
derived values given by Collier Cameron et al.~(2010).
Starred systems ($\star$) were
omitted from the sample discussed in \S\S~2-3.}

\end{deluxetable}

\end{document}